\RequirePackage[l2tabu, orthodox]{nag}

\documentclass[utf8]{lni}

\usepackage{booktabs}

\usepackage[math]{blindtext}

\usepackage{graphicx}

\usepackage{fixmetodonotes}

\usepackage[caption=false]{subfig}

\startpage{1}

\editor{Maximilian Eibl, Martin Gaedke}
\booktitle{INFORMATIK 2017}

\author[Andreas Arzt \and Matthias Dorfer]{
Andreas Arzt\footnote{Austrian Research Institute for Artitifical Intelligence, \email{andreas.arzt@ofai.at}} \and
Matthias Dorfer\footnote{Johannes Kepler University Linz, Department of Computational Perception, \email{matthias.dorfer@jku.at}}}

\title[Aktuelle Entwicklungen in der Automatischen Musikverfolgung]{Aktuelle Entwicklungen in der Automatischen Musikverfolgung}

\begin{document}
\maketitle

%Auf Anzahl der AutorInnen setzen, damit die weitere Nummerierung der Fußnoten passt
%Dieser Befehl \verb|\setcounter{footnote}{2}| legt in dem Fall fest, dass 2 Fußnotennummern bereits für die AutorInnen verbraucht wurden, damit die darauf folgenden Fußnoten mit der richtigen Nummerierung (ab 3) fortfahren. Dieser Wert muss an die jeweilige Zahl an AutorInnen bzw. bereits verbrauchte Fußnoten angepasst werden, sofern im weiteren Verlauf Fußnoten verwendet werden.
\setcounter{footnote}{2}

\begin{abstract}
Diese Arbeit befasst sich mit aktuellen Entwicklungen in der automatischen Musikverfolgung durch den Computer. Dieser Prozess ist auch unter den Begriffen "`Score Following"' oder "`Real-time Music Tracking"' bekannt. Es handelt sich dabei um Algorithmen, die einer musikalischen Aufführung "`zuhören"', das aufgenommene Audiosignal mit einer (abstrakten) Repräsentation des Notentextes vergleichen und sozusagen in diesem mitlesen. Der Algorithmus kennt also zu jedem Zeitpunkt die Position der Musiker im Notentext. Diese Information erlaubt die Realisierung einer Reihe von Anwendungen, zum Beispiel der automatischen Musikvisualisierung und der automatischen Begleitung. Neben der Vermittlung eines generellen Überblicks, liegt der Schwerpunkt dieser Arbeit auf der Beleuchtung des Aspekts der Flexibilität und der einfacheren Nutzbarkeit dieser Algorithmen. Es wird dargelegt, welche Schritte getätigt wurden (und aktuell getätigt werden) um den Prozess der automatischen Musikverfolgung einfacher zugänglich zu machen. Dies umfasst Arbeiten zur automatischen Identifikation von gespielten Stücken und deren flexible Verfolgung ebenso wie aktuelle Ansätze mithilfe von Deep Learning, die es erlauben Bild und Ton direkt zu verbinden, ohne Umwege über abstrakte und nur unter großem Zeitaufwand zu erstellende Zwischenrepräsentationen.

\end{abstract}


\section{Einleitung}

Diese Arbeit beschäftigt sich mit der automatischen Verfolgung von Musik durch Computeralgorithmen -- ein Prozess, der auch unter den Begriffen "`Score Following"' und "`Music Tracking"' bekannt ist. Das Ziel dieses Prozesses ist es, eine Aufführung von (vornehmlich klassischer) Musik mit dem zugrundeliegenden Notentext abzugleichen und sozusagen in diesem live mitzulesen.

Es handelt sich dabei also um Algorithmen, die einer musikalischen Aufführung "`zuhören"', das Audiosignal verarbeiten, es mit einer abstrakten Repräsentation des Notentextes vergleichen und diese beiden Zeitreihen (das Audiosignal und den Notentext) miteinander synchronisieren ("`alignen"'). Der Algorithmus berechnet zu jedem Zeitpunkt die Position der Musiker im Notentext. Mithilfe dieser Information lassen sich eine Reihe von Anwendungen realisieren, zum Beispiel automatische Begleitsysteme und die automatische Synchronisation von Medien wie Bild und Text zu der live gespielten Musik für künstlerische Zwecke.

Die ersten Arbeiten zur automatischen Musikverfolgung wurden in den 1980er Jahren veröffentlicht (siehe \cite{dannenberg:icmc:1984,vercoe:icmc:1984}). Diese Ansätze waren der Idee eines automatischen Begleitsystems verbunden und basierten noch auf symbolischem Input (sowohl der Notentext als auch die musikalische Aufführung werden als Zeichenketten repräsentiert) und Stringmatchingalgorithmen, um diese beiden Informationen zu synchronisieren. Die Notwendigkeit von symbolischen Daten der musikalischen Aufführung war eine große Einschränkung, da diese Algorithmen somit nur mit speziellen Instrumenten verwendbar waren. Deshalb trat diese Art des Verfolgens von Musik etwas in den Hintergrund (Ausnahmen sind beispielsweise \cite{nakamura:ismir:2015,xia:phd:2016}). Stattdessen liegt der Fokus aktuell hauptsächlich auf Systemen, die direkt auf dem Audiosignal der musikalischen Aufführung das Problem der automatischen Musikverfolgung zu lösen versuchen.

Die meisten aktuellen Verfahren basieren auf diversen Variationen von probabilistischen Modellen, beispielsweise auf Hidden Markov Modellen \cite{raphael:icml:2010,raphael:nips:2001,raphael:smc:2009,orio:icmc:2001,orio:nime:2003,cont:icassp:2006} und verwandten Modellen \cite{cont:tpami:2010,duan:icassp:2011,bochen:ismir:2015}, Conditional Random Fields \cite{sako:amt:2014, yamamoto:icassp:2013} und Partikelfiltern \cite{korzeniowski:icmc:2013, montecchio:icassp:2011,otsuka:eurasip:2011}. Eine Ausnahme stellen Musikverfolgungssysteme basierend auf Dynamic Time Warping (DTW) dar \cite{arzt:ecai:2008,arzt:phd:2016}.

All diesen Systemen ist gemein, dass sie relativ inflexibel sind und die Aufbereitung der für die automatische Verfolgung notwendigen Daten zeitaufwändig ist. Wir werden diese Problematiken in Kapitel \ref{sec:flex} detaillierter diskutieren. Daraufhin werden in den Kapiteln \ref{sec:anytime} und \ref{sec:end-to-end} zwei aktuelle Entwicklungen beschrieben, die auf einfachere und flexiblere Verwendung von Musikverfolgungsalgorithmen abzielen. Schließlich werden wir in Kapitel \ref{sec:conclusion} den aktuellen Stand der Forschung nochmals kurz diskutieren und einen Ausblick auf kommende Herausforderungen geben.

\section{Präzision vs. Flexibilität in der Automatischen Musikverfolgung}\label{sec:flex}

In den letzten Jahren lassen sich zwei Trends im Bereich der Musikverfolgung identifizieren, die auch stark von den jeweiligen Anwendungsgebieten der Algorithmen beeinflusst werden. Einerseits ist dies die Anwendung des \emph{automatischen Begleitens}, und andererseits die \emph{automatische Synchronisation von Visualisierungen}.

\emph{Automatische Begleitung} hat zum Ziel zur Aufführung eines Solomusikers automatisch Begleitelemente (einzelne Effekten oder die Simulation eines kompletten Orchesters) zu synchronisieren (siehe zum Beispiel \cite{cont:tpami:2010,raphael:smc:2009}). In diesem Fall ist das Ziel Präzision, beziehungsweise sogar die Vorhersage von Ereignisse, da selbst kleine Abweichungen sich als hörbare Fehler in der Aufführung manifestieren würden.

Um die notwendige Präzision für automatische Begleitung zu erzielen, ist die adäquate Aufbereitung des Notentextes unumgänglich. Dieser muss in einer symbolischen Repräsentation vorliegen, die es erlaubt die erwartete Aufführung des Stückes so exakt wie möglich zu modellieren. Dies umfasst  die Definition der Struktur (beispielsweise die genaue Festlegung, welche Wiederholungen gespielt oder ausgelassen werden) ebenso wie die Modellierung von komplexeren Konzepten wie Trillern oder Glissandi, deren spezifische Ausführung nicht explizit durch den Notentext definiert wird (beispielsweise ist die Anzahl der Noten, die ein Triller umfasst, nicht per se festgelegt). In einigen Fällen ist die Erstellung der symbolischen Beschreibung direkt Teil des Kompositionsprozesses -- die Komposition wird also bereits in Kombination mit dem spezifischen Musikverfolgungsalgorithmus erarbeitet \cite{coffy:smc:2014}.

Diese Modelle werden speziell auf die geplante Aufführung abgestimmt. Dies geschieht normalerweise indem die Parameter des Modells auf einer Reihe von Probeaufführungen, die möglichst ähnlich der tatsächlichen Aufführung sein sollten, optimiert werden. Ein automatisches Begleitsystem für ein bestimmtes Stück beziehungsweise eine spezielle Aufführung vorzubereiten ist also ein komplexer und zeitaufwändiger Prozess.

\begin{figure}[tb]
 \centerline{
     \includegraphics[height=3.2cm]{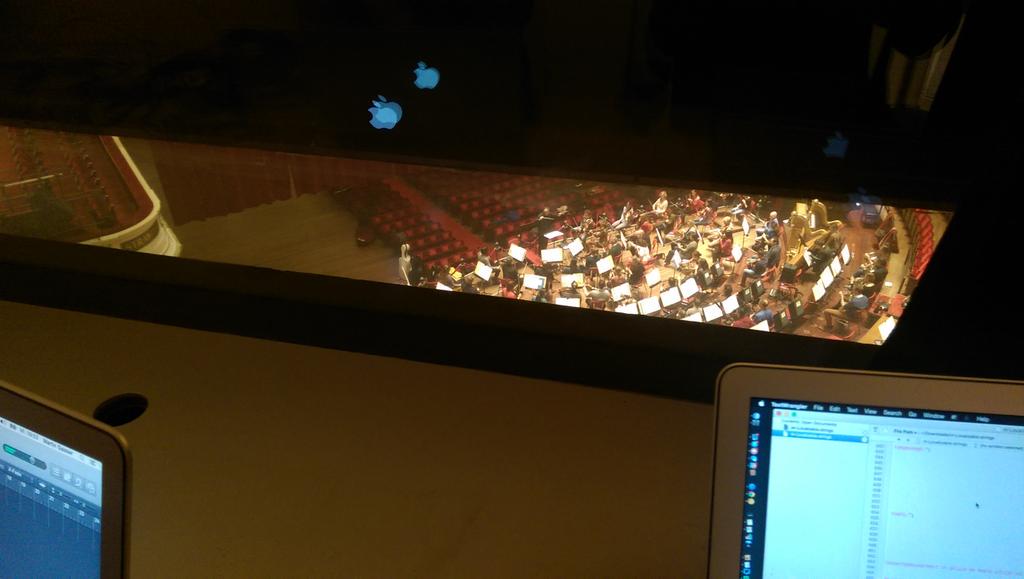}
     \hspace{0.5cm}
     \includegraphics[height=3.2cm]{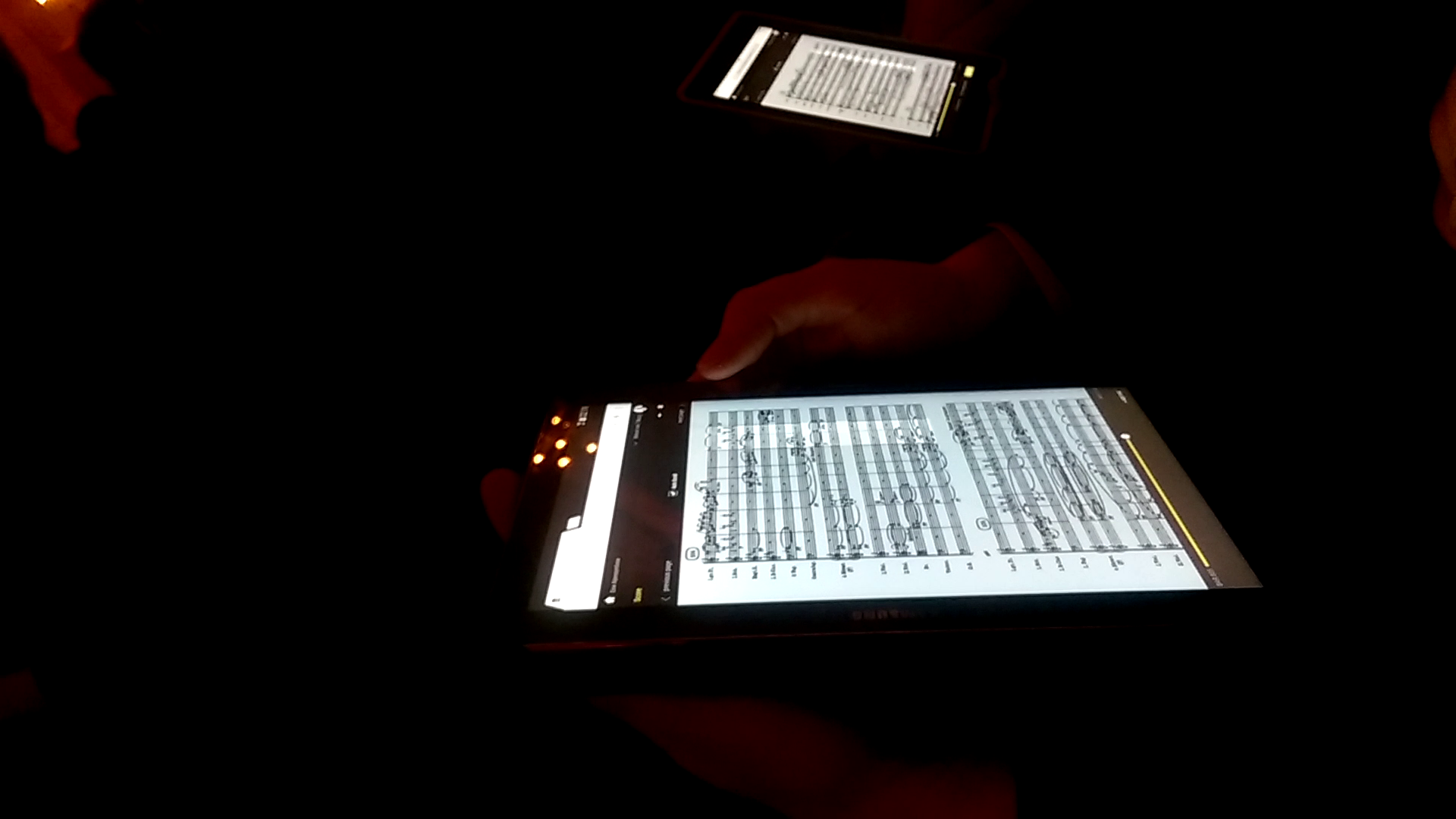}
    }
 \caption{Impressionen eines Experiments im Rahmen eines regulären Konzertes im Concertgebouw in Amsterdam. Links: Blick vom Kontrollraum auf die Bühne (während einer Orchesterprobe); Rechts: Zur Musik synchronisierter Notentext, angezeigt auf persönlichen Tabletcomputern im Publikum (aus \cite{arzt:ijcai:2015}).
 \label{fig:livepictures_concertgebouw}}
\end{figure}

Wie bereits erwähnt, ist das zweite große Anwendungsgebiet die \emph{automatische Synchronisation von Visualisierungen} zur Musik. Dies umfasst beispielsweise die Anzeige eines zur Musik synchronisierten Notentextes (inklusive automatischem Umblättern) ebenso wie Medien wie Text, Bild oder Video, die dem Publikum im richtigen Moment präsentiert werden, um einen künstlerischen Mehrwert zu generieren oder einen weiterbildenden Effekt zu erzielen (siehe Abbildung \ref{fig:livepictures_concertgebouw}, die Impressionen eines Tests einer solchen Anwendung im Concertgebouw in Amsterdam zeigt \cite{arzt:ijcai:2015}). Hier ist Präzision zwar noch immer eine wünschenswerte Eigenschaft, aber nicht notwendigerweise im selben Ausmaße. Um zum Beispiel automatisch den Notentext synchronisiert zur Musik anzuzeigen und umzublättern sind normalerweise Abweichungen im Bereich von einigen hundert Millisekunden durchaus akzeptabel.

Während also ein gewisses Maß an Genauigkeit natürlich weiter erforderlich ist, treten andere Aspekte ebenso in den Vordergrund. Von generellem Interesse ist beispielsweise die Reduktion des Aufwandes, der notwendig ist, um einen Notentext für die automatische Musikverfolgung aufzubereiten. Ein wichtiger Aspekt in vielen Anwendungen ist auch die Flexibilität im Umgang mit der Aufführung an sich. Nützlich kann beispielsweise ein Algorithmus sein, der "`ungeplante"' Abweichungen des Musikers vom Notentext (wie das Auslassen einer Wiederholung) erkennt und dynamisch darauf korrekt und so schnell wie möglich reagiert. In dieser Arbeit werden wir nun das Streben nach Präzision, und damit die Arbeit an automatischen Begleitsystemen, außer Acht lassen und stattdessen zwei aktuelle Forschungsrichtungen zum flexiblen Musikverfolgen genauer betrachten. 

Die erste Forschungsrichtung, beschrieben in Kapitel \ref{sec:anytime}, beschäftigt sich mit der Frage nach Flexibilität während einer musikalischen Performance. Wir werden hier einen Ansatz zusammenfassen, der in der Lage ist innerhalb weniger Sekunden Musikstücke zu identifizieren. Damit wird ein Musikverfolgungssystem realisiert, das flexibles Verfolgen auf Basis einer großen Datenbank an Notentexten erlaubt.

Die zweite Forschungsrichtung, präsentiert in Kapitel \ref{sec:end-to-end}, setzt noch einen Schritt früher an und versucht den Vorverarbeitungsprozess, der klassischerweise notwendig ist um den Notentext für automatische Musikverfolgung aufzubereiten, obsolet zu machen. Dazu werden Verfahren des maschinellen Lernens verwendet, um direkt Korrespondenzen zwischen Bilddaten und Audiosignalen zu lernen. Die präsentierten Ansätze stecken zwar noch in den Kinderschuhen, sind aber erste wichtig Schritte hin zu unserer Vision von einfacher Nutzung von automatischer Musikverfolgung für jedermann: eine Anwendung auf einem (mobilen) Endgerät, die anhand von Fotos von Seiten von Notentexten in der Lage ist, ohne jedweden manuellen Eingriff, eine Aufführung zu verfolgen.

\section{Ein flexibles Musikverfolgungssystem basierend auf einer Datenbank symbolischer Notentexte}\label{sec:anytime}

Der Prozess des automatischen Musikverfolgens ist normalerweise relativ starr definiert. Es wird ein definiertes Stück von Anfang bis Ende verfolgt, wobei keine strukturellen Abweichungen erlaubt sind. Ein flexiblerer Ansatz wird in \cite{arzt:phd:2016} vorgestellt.

Dieses System basiert auf einer Datenbank von symbolischen Notentexten klassischer Klaviermusik und enthält unter anderem alle Sonaten Mozarts und Beethovens, sowie Chopins Gesamtwerk für Soloklavier (insgesamt mehr als eine Million Noten). Es erlaubt flexible Musikverfolgung auf der gesamten Datenbank von Stücken und kann automatisch das Stück, sowie die exakte Position innerhalb des Stückes erkennen und dann der Aufführung folgen. Ebenso kann es beliebige Sprünge innerhalb eines Stückes und auch zu anderen Stücken detektieren und die korrekte Position wiederfinden. Es kann also jeder Aktion des Musikers / der Musikerin innerhalb dieser Datenbank von Notentexten folgen.

\begin{figure}[tp]
\begin{center}
\includegraphics[width=10cm]{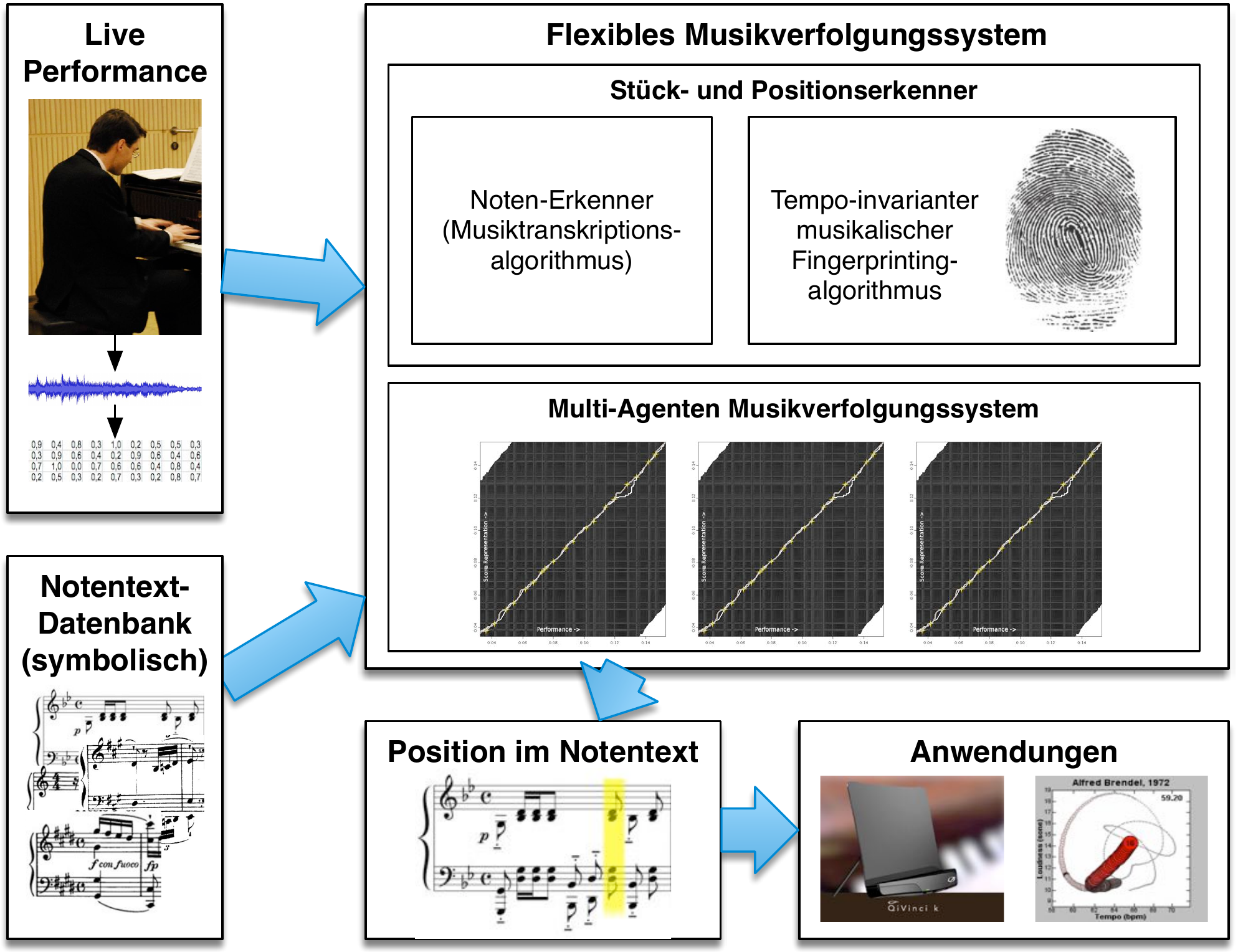}
\caption{Schematische Darstellung des flexiblen Musikverfolgungssystems (adaptiert von \cite{arzt:phd:2016}).}
\label{fig:tracker}
\end{center}
\end{figure}

Abbildung \ref{fig:tracker} skizziert die Funktionsweise dieses Systems. Das Erkennen der Position über alle in der Datenbank vorhandenen Notentexte wird durch eine Kombination aus einem Musiktranskriptionsalgorithmus \cite{boeck:icassp:2012} (dieser Algorithmus ist auf Klaviermusik trainiert und schränkt somit die Anwendung des Systems auf ebendiese ein) und einem tempo-invarianten musikalischen Fingerprinting-Algorithmus \cite{arzt:ismir:2012} realisiert. Zuerst übersetzt der Transkriptionsalgorithmus das Audiosignal der Aufführung in eine Sequenz symbolischer Information (eine Liste von Tonhöhen inklusive deren Beginnzeiten). Mithilfe des Fingerprinting-Algorithmus wird die Datenbank von symbolischen Notentexten nach zum aktuellen Kontext in der Aufführung möglichst ähnlichen Ausschnitten durchsucht. Dieser Prozess läuft dauerhaft im Hintergrund und berechnet Hypothesen, die durch das Multiagenten-Musikverfolgungssystem weiterverarbeitet werden.

Das Multiagenten-Musikverfolgungssystem basiert auf mehreren Instanzen eines Musikverfolgungssystems, das Audio-zu-Audio Alignment mithilfe einer Echtzeitversion des Dynamic Time Warping Algorithmus realisiert \cite{dixon:ijcai:2005,arzt:ecai:2008,arzt:phd:2016}. Dieses übernimmt die Positionshypothesen und versucht die musikalische Aufführung anhand dieser Notentextpositionen mitzuverfolgen. Falls es sich bei einer Hypothese um die korrekte Position handelt, wird dies zu geringen Alignmentkosten führen, während der Versuch dies anhand einer inkorrekten Position zu tun zu hohen Kosten führen wird. Zu jedem Zeitpunkt wird die Instanz des Verfolgungsalgorithmus, die die geringsten Alignmentkosten verursacht, als aktiv markiert -- sie repräsentiert also die aktuelle Positionshypothese des Systems über alle Stücke in der Datenbank.

Dieses System erlaubt flexibles automatisches Musikverfolgen anhand einer großen Sammlung von (symbolischen) Notentexten klassischer Klaviermusik. Neben Anwendungen für Musiker, wie unterstützende Übungssysteme, bietet es sich an, diese Algorithmen als Grundgerüst für mobile Anwendungen für Konsumenten klassischer Musik zu nutzen. Beispielsweise ließe sich eine App realisieren, die dem Liebhaber klassischer Klaviermusik vollkommen autonom Zusatzinformationen zur Musik liefert. Diese Anwendung könnte automatisch erkennen welches Stück gespielt wird und den Notentext und künstlerische oder weiterbildende Visualisierungen (etwa die Struktur des Stückes oder Informationen über wichtige musikalische Themen) einblenden. Falls das Stück entsprechend aufbereitet wurde, könnten historische Hintergrundinformation über den Komponisten und das Stück bereitgestellt werden. Ebenso könnte auf weitere berühmte Interpretation des Stücks (und deren Bezugsquellen) hingewiesen werden.

\section{Automatische Musikverfolgung direkt anhand des graphischen Notentextes}\label{sec:end-to-end}

Ein großes Hindernis, das einer weiter verbreiteten Verwendung von automatischer Musikverfolgung im Weg steht, ist der immense Zeitaufwand der notwendig ist, um die Notentexte für den Verfolgungsprozess aufzubereiten. Der größte Aufwand wird durch die Erstellung beziehungsweise Aufbereitung einer symbolischen Beschreibung des Notentext (zum Beispiel MusicXML oder auch MIDI) verursacht. Diese Beschreibung ist notwendig um den Notentext exakt modellieren zu können.

Es gibt zwar automatische Verfahren via sogenannter Optical Music Recognition (OMR) Algorithmen um aus dem graphischen Notentext diese symbolische Repräsentation zu extrahieren, diese sind aber momentan nur sehr eingeschränkt für komplexere Notentexte geeignet \cite{wen2015omr,hajic:ismir:2016,byrd:jnmr:2015,rebelo:jmir:2012}. Ein Versuch den Notentext für das zur automatischen Musikverfolgung verwandte Gebiet des Audio Alignment via OMR aufzubereiten, ist in \cite{kurth2007automated} beschrieben.

% In this section we summarize and outline a novel alignment paradigm trying to perform score following directly in images of sheet music.
% Before we describe our proposals in more detail we first review the most related work in this field.
% Initial work trying to solve score following directly in sheet music images mainly relies on an intermediate optical music recognition steps (cite Meinard, ...).
% In particular, they first analyze the scores (sheet music) with the purpose to extract so called score-chroma features.
% As a next step they also extract chroma features from the audio.
% Given this sheared chroma mid-level representation it is possible to perform score following by dynamic time warping on score-audio-chroma distance matrices.

In diesem Kapitel wird ein neuer Ansatz vorgestellt, der komplett auf jedwede Zwischenrepräsentation verzichtet und direkt auf dem graphischen Notentext arbeitet. Der Ansatz basiert auf aktuellen Forschungsergebnissen aus dem Bereich der künstlichen Intelligenz, genauer gesagt auf künstlichen neuronalen Netzen,
welche auch unter dem Schlagwort "`Deep Learning"' populär geworden sind. Die angesprochenen Fortschritte im maschinellen Lernen machen es möglich Korrespondenzen zwischen Audiodaten (der musikalischen Aufführung) und Bilddaten (dem Notentext) \emph{direkt} zu lernen. Diese Algorithmen lernen also gleichzeitig einen Notentext zu "`lesen"', Musik zu "`hören"', und das Gelesene und Gehörte miteinander in Beziehung zu setzen. Die neuronalen Netze erlernen diesen Vorgang rein durch Training anhand von Beispielen.

% In the following, we propose a new class of models which all try to circumvent intermediate representations e.g. using OMR to extract score-chroma features.
% Our methods are based on artificial neural networks which aim at linking the audio and the image of the sheet music directly, by learning correspondences between these two modalities,
% and thus making the complicated step of creating an in-between representation obsolete.
% In particular, we try to develop algorithms that simultaneously learn to read notes,
% listen to music and match the currently played music with the correct notes in the sheet music.
% To be clear, the networks learns their behaviour solely from the examples shown for training.

Nachfolgend werden wir zwei unterschiedliche,  komplementäre Ansätze vorstellen, die beide dieses Paradigma verfolgen: Automatische Musikverfolgung via Audio-zu-Notentext Matching (siehe Kapitel \ref{sec:tracking_md_1}) und Automatische Musikverfolgung via Multi-Modalen Joint Embedding Spaces (siehe Kapitel \ref{sec:tracking_md_2}).

Beide Arbeiten sind noch frühe Prototypen und wurden bisher hauptsächlich mit aus MIDI Dateien synthetisierten Audiodaten (via Klaviersoundfonts) trainiert und getestet. Nichtsdestotrotz handelt es sich um vielversprechende Ansätze, die in einigen Jahren automatische Musikverfolgung für jedermann zugänglich machen könnten. Anstelle einer komplizierten Datenaufbereitung würde es ausreichen direkt auf einem mobilen Endgerät den Notentext zu fotografieren. Ohne weitere manuelle Eingriffe könnte danach das Musikverfolgen gestartet werden.

% By now we have discovered two different, but complementary, learning based multi-modal neural networks:

% \begin{enumerate}
% \item Score-Following by Audio to Sheet Music Matching
% \item Score-Following by Multi-Modal Joint Embedding Spaces (Score-Following by Cross-Modality Retrieval)
% \end{enumerate}

\subsection{Automatische Musikverfolgung via Audio-zu-Notentext Matching}\label{sec:tracking_md_1}

\begin{figure}[tb]
 \centering
 \includegraphics[width=.6\textwidth]{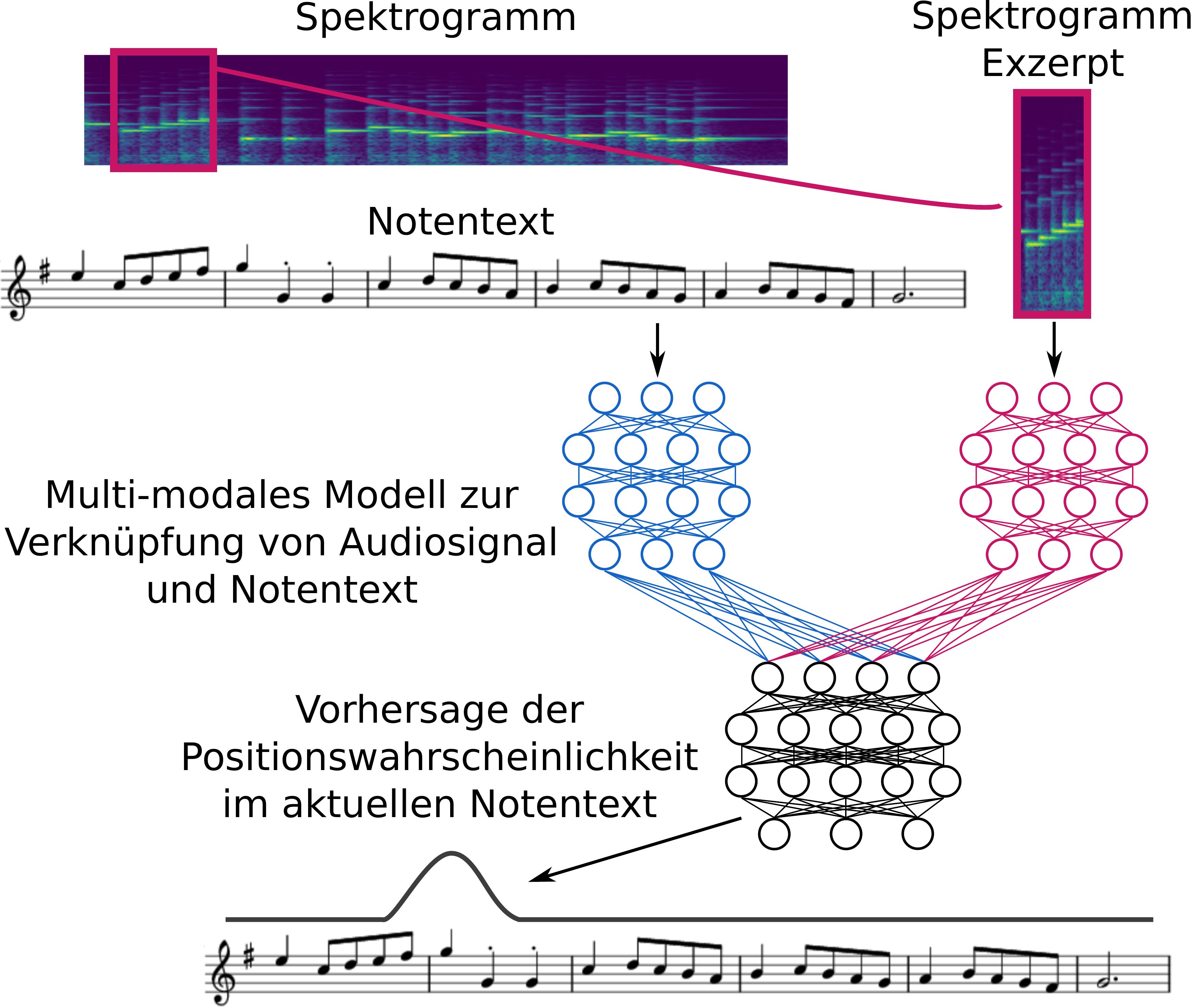}
 \caption{Schematische Darstellung des automatischen Musikverfolgungsprozesses via Audio-zu-Notentext Matching. Die beiden oberen, getrennten Teile des multi-modalen Netzes verarbeiten die jeweilige Modalität (Notentext und Audio). Der untere Teil des Netzes führt beide Modalitäten zusammen und sagt die Positionswahrscheinlichkeit im aktuellen Notentext vorher.}
 \label{fig:task_matching_probability}
\end{figure}

Die Idee dieses Ansatzes ist gleichzeitig den aktuellen Kontext im Audiosignal und ein Bild des Notentextes zu verarbeiten und in Beziehung zueinander zu setzen. In \cite{Dorfer2016Towards} wurden erste Versuche auf sehr reduzierter Klaviermusik (monophoner Notentext und monophones Audiosignal) beschrieben. Ein Schema dieses Ansatzes ist in Abbildung \ref{fig:task_matching_probability} zu sehen. Da eine genaue Beschreibung des Algorithmus den Rahmen dieser Arbeit sprengen würde, skizzieren wir nachfolgend seine generelle Funktionsweise.

% \emph{Score-Following by Audio to Sheet Music Matching} takes both sheet music and audio as input at the same time to predict which location in the sheet image best matches the current audio excerpt.
% [Dorfer et al., 2016] have shown in their initial experiments on simple piano music
% how this model design can be emploeyd to perform score following directly in images of sheet music.
% A sketch of this working pricipal is shown in Figure \ref{fig:task_sketch_matching}.

% %
% \begin{figure}
%  \centering
%  \includegraphics[width=.8\textwidth]{figs/task_sketch_german.pdf}
%  \caption{Score following by audio-to-sheet-music matching.}
%  \label{fig:task_sketch_matching}
% \end{figure}
% %

Das zentrale Element dieses Ansatzes sind zwei parallele Convolutional Neural Networks, je eines für die Audioeingabedaten beziehungsweise den Notentext. Diese beiden Netze lernen abstrakte interne Repräsentationen der jeweiligen Modalität die im Anschluss durch einen gemeinsam Layer miteinander verbunden werden. Das Ziel dieses multimodalen Netzwerks ist es, eine Wahrscheinlichkeitsverteilung abzuschätzen, die den aktuellen Audiokontext in Relation zum Notentext setzt -- also wiedergibt, wie wahrscheinlich der gegebene Ausschnitt an Audiodaten den jeweiligen Positionen im Notentext entspricht.
Gegeben diese Verteilung kann die aktuelle Position durch finden des Maximums dieser Verteilung berechnet werden.

% As a detailed description of the underlying methodology would be beyound the scope of this paper
% we summarize the general working principal.
% The heart of the method is a multi-modal neural network comprising two different convolutional pathways.
% One for processing the sheet image and one for processing the audio.
% Each of the two pathways learns an abstract, internal representation of the respective modality,
% which is than fused by a sheared multi-modal neural network.
% The objective of this shared multi-modal neural network is to predict a probability distribution
% over the most likely position in the given score images with respect to the current audio excerpt.
% Given this probability distribution we can use it for deriving the current location in the score by simply
% looking for the region having the highest probability density (see Figure \ref{fig:task_matching_probability}).

\subsection{Automatische Musikverfolgung via Multi-Modalen Joint Embedding Spaces}\label{sec:tracking_md_2}

Der zweite auf multi-modalen neuronalen Netzen basierende Ansatz ist enger mit der flexiblen Musikverfolgung, wie in Kapitel \ref{sec:flex} vorgestellt, verwandt. Hier wird das Problem des Findens von Korrespondenzen zwischen Audiodaten und Bilddaten vom Blickpunkt einer Retrievalaufgabe zwischen den beiden Modalitäten betrachtet \cite{dorfer:ismir:2017}.

\begin{figure}[t]
 \centering
 \includegraphics[width=.4\textwidth]{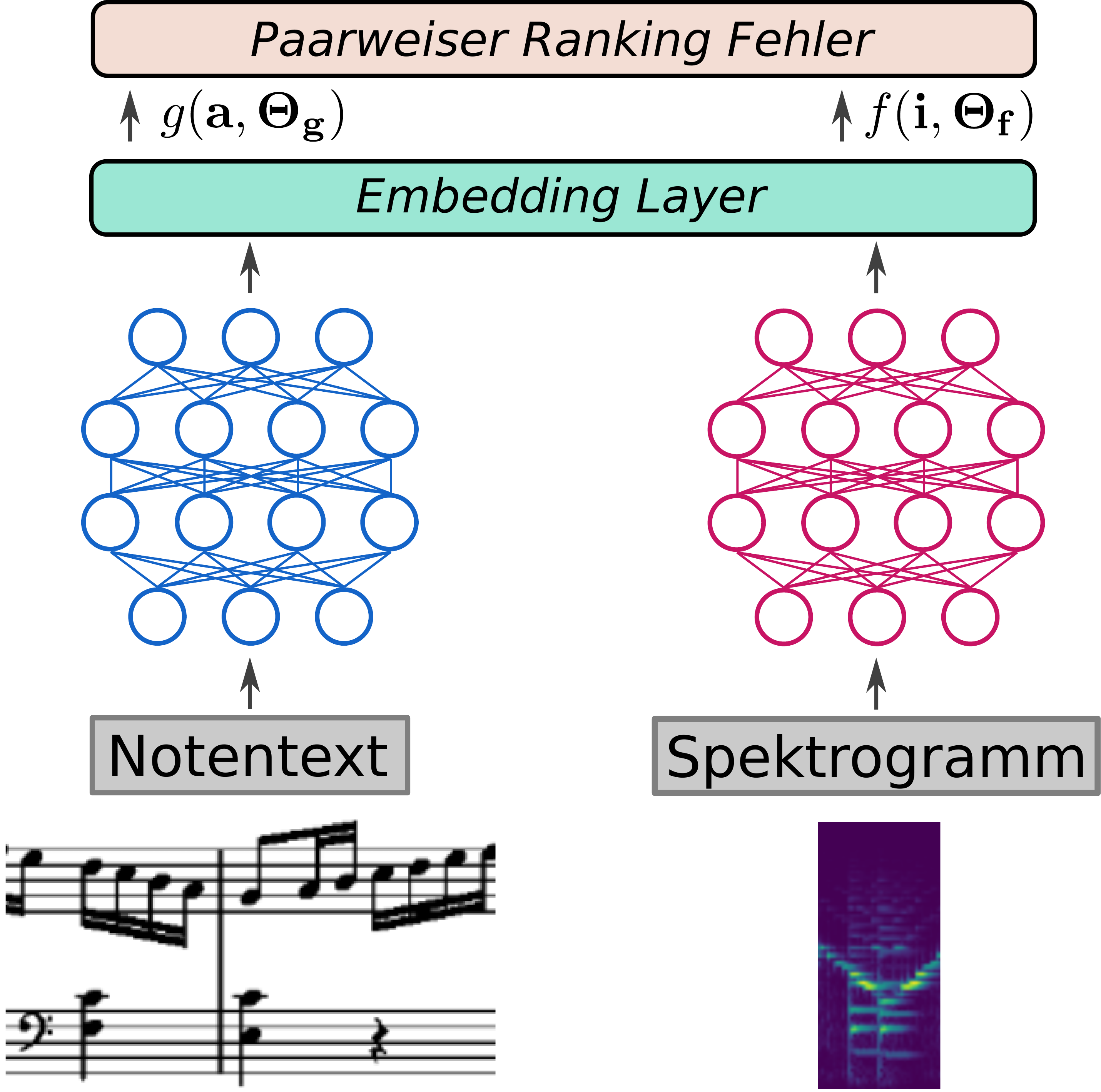}
 \caption{Architektur des Retrievalnetzwerks. Das Netzwerk wird mit dem Ziel die Ähnlichkeit von korrespondierenden Paaren von Audio- und Notentextdaten zu maximieren trainiert. Dies wird durch das Minimieren eines paarweisen Rankingfehlers erreicht.}
 % Architecture of correspondence learning network. The network is trained to optimize the similarity (in embedding space) between corresponding audio and sheet image snippets by minimizing a pair-wise ranking loss. 
 \label{fig:retrieval_network}
\end{figure}

Während des Trainings werden dem neuronalen Netz wieder Paare von kurzen Audiodaten und Ausschnitten von Bildern von Notentexten präsentiert.
Allerdings wird in diesem Ansatz versucht beide Modalitäten in einem gemeinsamen Einbettungsraum ("`Embedding Space"') zu repräsentieren.
Dazu werden während des Trainings für beide Modalitäten Projektionen (in unserem Fall nichtlineare Projektionen mittels neuronaler Netze) mit folgender Eigenschaft gelernt:
Die Projektionen (Einbettungskoordinaten) zusammengehöriger Audio- und Bilddaten sollen kleine Distanzen zueinander aufweisen, wohingegen die Projektionen nicht korrespondierender Paare möglichst weit auseinander liegen sollen.
Abbildung \ref{fig:retrieval_network} skizziert die Architektur dieses Retrievalnetzwerks wobei $f(\mathbf{a}, \Theta_f)$ und $g(\mathbf{i}, \Theta_g)$ die Projektionen
von Audio $\mathbf{a}$ und Bildausschnitt $\mathbf{i}$ in den Einbettungsraum bezeichnen.
$\Theta_f$ und $\Theta_g$ sind die Parameter der jeweiligen Netze.
Um die oben beschriebenen Lagebeziehungen zwischen den Projektionen von korrespondierenden und nicht korrespondierenden Paaren zu erreichen, werden die Netzwerke dahingehenden optimiert, einen auf Kosinusdistanz basierenden \emph{paarweisen Rankingfehler} zu minimieren (für eine detaillierte Beschreibung dieses Optimierungsvorgangs verweisen wir auf \cite{dorfer:ismir:2017}).

\begin{figure}[t]
 \centering
 \includegraphics[width=.7\textwidth]{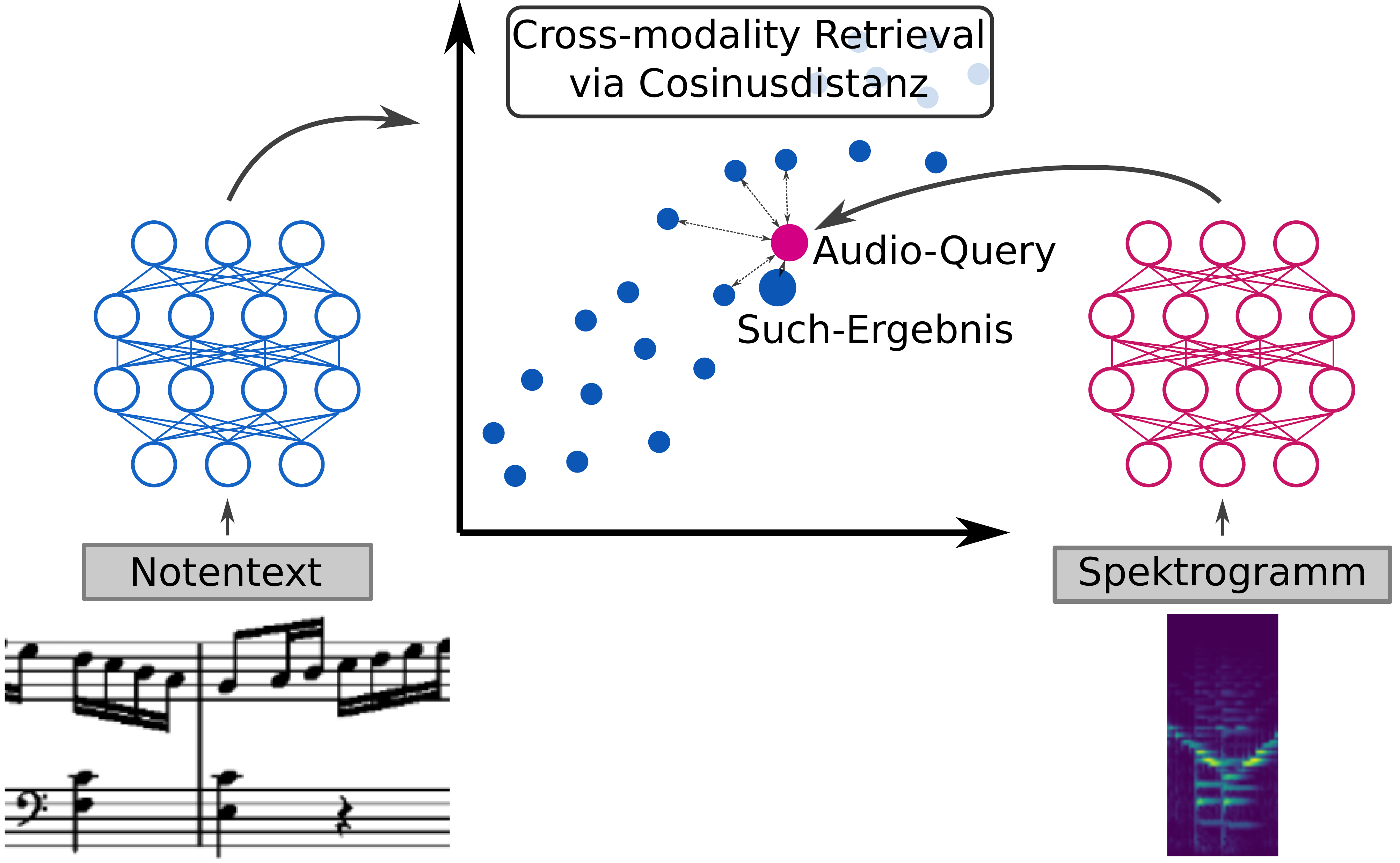}
 \caption{Der Retrievalprozess. Blaue Punkte repräsentieren eingebettete Notentextinstanzen. Der rote Punkt symbolisiert die Projektion der Audioabfrage in ebendiesen Raum. Der große blaue Punkt ist die Notentextinstanz, die der eingebetteten Audioabfrage am nächsten ist.}
%Sketch of sheet-music-from-audio retrieval. The blue dots represent the embedded candidate sheet music snippets. The red dot is the embedding of an audio query. The larger blue dot highlights the closest sheet music snippet candidate selected as retrieval result. 
\label{fig:retrieval_process}
\end{figure}

Nach dem Training können beide Projektionen unabhängig voneinander verwendet werden. Zum Beispiel kann zuerst eine Referenzdatenbank bestehend aus Bildern von Notentexten mithilfe der gelernten Projektion $f$ für Notentextbilder in diesen Einbettungsraum projiziert werden. Danach kann eine Audioabfrage mithilfe der Projektion $g$ für die Audiodaten in denselben Raum transferiert werden.
Auf Basis der oben beschriebenen Distanz (bzw. Lage) Eigenschaften kann der Raum nun mittels Kosinusdistanz nach Notentextausschnitten, die der Audioabfrage ähnlich sind, durchsucht werden.
Abbildung \ref{fig:retrieval_process} skizziert diesen Vorgang.

Dieses Szenario ist in der Literatur auch als Cross-Modality Retrieval bekannt. Um damit automatische Musikverfolgung zu realisieren, wird zuerst der Notentext (oder, wie in der flexiblen Musikverfolgen in Kapitel \ref{sec:flex} eine Sammlung von Notentexten) in den Suchraum eingebettet. Während der Aufführung wird der aktuelle Audiokontext laufend als Abfrage verwendet.
Das tatsächliche Musikverfolgen findet dann durch die zeitliche Analyse der Abfrageergebnisse analog zu traditionellen Ansätzen (etwa mithilfe von DTW) statt.

% (Abbildung adaptiert von \cite{dorfer:ismir:2017})
% (adaptiert von \cite{dorfer:ismir:2017})

% \emph{Score-Following by Multi-Modal Joint Embedding Spaces} addresses the problem from a cross-modality retrieval point of view.
% In contrast to the method described above this class of models addresses a more general scenario where both input modalities are required only at training time, for learning the relation between score and audio. This in turn requires a different network architecture that can learn two separate projections,
% one for embedding the sheet music and one for embedding the audio.
% Once the network is trained the two projections can then be used independently of each other.
% For example, we can first embed a reference collection of sheet music images using the image embedding
% part of the network, then embed a query audio and search for its nearest sheet music neighbours in the joint embedding space.
% This general scenario is referred to as cross-modality retrieval and supports different applications.

% Figure \ref{fig:retrieval_network} shows a sketch of such a retrieval network and Figure \ref{fig:retrieval_process} outlines the retrieval process from an audio excerpt query point of view.
%

\section{Zusammenfassung und Ausblick}\label{sec:conclusion}

Diese Arbeit beschäftigt sich mit flexiblen Algorithmen zur automatischen Musikverfolgung. Zum einen wurde eine Methode diskutiert, die flexibles Verfolgen basierend auf einer Datenbank von Notentexten erlaubt. Diese Methode ist bereits robust genug um komplexe Klaviermusik zu verfolgen und könnte als Grundgerüst für (mobile) Anwendungen verwendet werden. Zum anderen wurden in dieser Arbeit Ansätze diskutiert, die es erlauben automatische Musikverfolgung flexibler einzusetzen, indem die aufwändige explizite Erstellung einer symbolischen Repräsentation vermieden wird. Hierzu wurden zwei miteinander verwandte Methoden beschrieben, die direkte Korrespondenzen zwischen Bilddaten des Notentextes und Audiodaten einer Aufführung automatisch lernen. Dies bedeutet, dass die Qualität des gelernten Modells stark von der Anzahl und Qualität der verfügbaren Trainingsdaten abhängt. Die Erweiterung dieses Trainingsdatensatzes hat also hohe Priorität um noch generellere Modelle zur Verbindung von Audio- und Bilddaten zu trainieren. Das Hauptziel dieser Vergrößerung und Diversifizierung des Trainingsdatensatzes ist es diesen Ansatz auch mit Audiodaten von tatsächlichen Aufführungen klassischer Klaviermusik (beziehungsweise zu einem späteren Zeitpunkt beliebiger klassischer Musik) verwenden zu können -- bisher wurden diese Modelle wie erwähnt mit aus MIDI generierten Audiodaten getestet. Dies ist nicht trivial, da die neuronalen Netze dazu lernen müssen mit Herausforderungen wie asynchronen Notenonsets innerhalb eines Akkords, die Verwendung der Pedale des Klavieres und variierende Dynamik entsprechend zu modellieren. Die bisherigen Ergebnisse sowie der potentielle Nutzen eines generellen notentextbasierten automatischen Musikverfolgungssystems deuten für uns jedoch darauf hin, dass es vielversprechend ist diese Forschungsrichtung verstärkt zu verfolgen.

\section*{Acknowledgements}

Diese Arbeit wurde von den österreichischen Bundesministerien BMVIT und BMWFW und dem Bundesland Oberösterreich (COMET Center SCCH) sowie von der Europäischen Union (European Research Council, ERC Grant Agreement 670035, project CON ESPRESSIONE) unterstützt. Die Tesla K40, die für diese Arbeit verwendet wurde, wurde von der NVIDIA Corporation bereitgestellt.

\bibliography{paper-bibtex}

\end{document}